\documentstyle[12pt]{article}
\def\beqar {\begin{eqnarray}}
\def\eeqar {\end{eqnarray}}
\def\beq {\begin{equation}}
\def\eeq {\end{equation}}

\def \half {{\textstyle{1\over 2}}}

\def \la {{\langle}}
\def \ra {{\rangle}}

\def \Tr {{\rm Tr}}

\begin{document}

\begin{titlepage}
\null\vspace{-62pt}

\pagestyle{empty}
\begin{center}
\rightline{}
\rightline{CCNY-HEP-00/7 }
\rightline{RU-00-12-B }
\rightline{hep-th/0011172 }

\vspace{1.0truein} {\large\bf Quantum Mechanics on the
Noncommutative Plane and Sphere }\\
\vskip .1in {\large\bf }

\vspace{1in} V. P. NAIR$^a$ and A.P. POLYCHRONAKOS$^{b,}$
\footnote{On leave from Theoretical Physics Dept., Uppsala University, Sweden
and Physics Dept., University of Ioannina, Greece.}\\
\vskip .1in {\it $^{a,b}$Physics Department, City College of
the CUNY\\ New York, NY 10031\\
\vskip .05in
$^{a,b}$The Graduate School and University Center\\ City
University of New York\\ New York, NY 10016\\
\vskip .05in
$^b$Physics Department, Rockefeller University\\ New York, NY 10021\\}
\vskip .05in {\rm E-mail: vpn@ajanta.sci.ccny.cuny.edu, 
poly@teorfys.uu.se}\\
\vspace{1.in}
\centerline{\bf Abstract}

\end{center}
\baselineskip=18pt

We consider the quantum mechanics of a particle on a
noncommutative plane. The case of a charged particle in a
magnetic field (the Landau problem) with a harmonic oscillator
potential is solved. There is a critical point with the
density of states becoming infinite for the value of the
magnetic field equal to the inverse of the noncommutativity
parameter. The Landau problem on the noncommutative
two-sphere is also solved and compared to the plane problem.

\end{titlepage}

\hoffset=0in
\newpage
\pagestyle{plain}
\setcounter{page}{2}
\newpage

\section{Introduction} Noncommutative spaces can arise as
brane configurations in string theory and in the matrix model
of M-theory \cite{bfss}. Fluctuations of branes are described
by gauge theories and thus, motivated by the existence of
noncommutative branes, there has  recently been  a large
number of papers dealing with gauge theories, and more
generally field theories, on such spaces \cite{everyone}.
However, there has been relatively little work exploring the
quantum mechanics of particles on noncommutative spaces.
Since the one-particle sector of field theories, which can be
treated in a more or less self-contained way in the free
field or weakly coupled limit, leads to quantum mechanics,
the brane connection suggests that a more detailed study of
this topic should be useful. This is the subject of the
present paper.

Some of the algebraic aspects of quantum mechanics on spaces
with an underlying Lie algebra structure were considered in
reference \cite{vpn}. The noncommutative plane can be defined 
in terms of a projection to the lowest Landau level of dynamics 
on the commuting plane \cite{DJT}; some features of particle
dynamics in terms of a similar construction were contained in 
\cite{bigatti}. 
The spectrum of a harmonic oscillator on the noncommutative
plane was derived in \cite{LSZ} and the case of a general 
central potential was recently  discussed in
\cite{gamboa}. In this paper, we will analyze the algebraic
structures in more detail. We will solve the problem of a
charged particle in a magnetic field (the Landau problem)
with an oscillator potential on the noncommutative plane.
There is an interesting interplay of the magnetic field $B$
and the noncommutativity parameter $\theta$, with a critical
point at
$B\theta =1$ where the density of states becomes infinite. We
also solve the Landau problem on the noncommutative sphere,
for which the basic algebraic structure turns out to be
$SU(2)\times SU(2)$. We also show how the results on the
plane can be recovered in the limit of a large radius for the
sphere.

\section{The noncommutative plane}

We start with the quantum mechanics of a particle on the
noncommutative two-dimensional plane. For single particle
quantum mechanics, we need the Heisenberg algebra for the
position and momentum operators. The two-dimensional
noncommutative plane is described by the coordinates $x_1,
x_2$ which obey the commutator algebra $[x_1, x_2] =i\theta$
where $\theta$ is the noncommutativity parameter. With the
momentum operator $p_i,~i=1,2$, we may write the full
Heisenberg algebra as
\beqar &&[x_1,x_2]= i\theta \nonumber\\ &&[x_i , p_j] ~=
i\delta_{ij}\label{1}\\  &&[p_1,p_2]~= 0 \nonumber
\eeqar The fact that $x_1$ and $x_2$ commute to a constant
may suggest that they can themselves serve as translation
operators. However, this is not adequate to obtain the last
of the relations (\ref{1}); one needs independent operators.
A realization of the momentum operators, for example, would be
\beqar &&p_1 ={1\over \theta} \left(~ x_2 +k_1
\right)\nonumber\\ &&p_2= {1\over \theta}\left(-x_1
+k_2\right)\label{2}
\eeqar with $[k_1,k_2]=-i \theta$ and $[k_i,x_j]=0$. In this
case, $(x_1,x_2)$ and $( k_2,  k_1)$ obey identical
commutation rules and are mutually commuting. $p_i$ are  thus
constructed from two copies of the $x$-algebra. 

We may use the realization (\ref{2}) of the $p_i$ to solve
specific quantum mechanical problems. However, before turning
to specific examples, some comments about the $p_i$-operators
are in order. In the usual quantum mechanics with commuting
$x$'s, a single irreducible representation for the
$x$-algebra would be given by $x_i =c_i$ for fixed real
numbers $c_i$. Coordinate space is spanned by an infinity of
irreducible representations of the $x$-algebra. Additional
independent operators
$p_i$ are needed to obtain a single irreducible
representation, now for the augmented set of operators. The
$p_i$'s connect different irreducible representations of the
$x$-algebra. In order to recover this structure for small
$\theta$, we need the independent set of operators $k_i$ in
(\ref{2}).

Single particle quantum mechanics may also be viewed as the
one-particle sector of quantum field theory, in the free
field or very weakly coupled limit, with the Schr\"odinger
wave function obeying essentially the free field equation.
Since quantum field theories on noncommutative manifolds have
already been defined and investigated to some extent, this
may seem to give a quick and simple way to write down
one-particle quantum mechanics. The case of a nonrelativistic
Schr\"odinger field suffices to illustrate the point. The
field
$\Phi (x)$ is a function of the noncommuting coordinates
$x_i$. The action for this field in an external potential and
coupled to a gauge field may be written as
\beq {\cal S} =\int dt ~\Tr \left[ {\Phi^\dagger iD_0 \Phi }
-{ (D_i\Phi)^\dagger (D_i \Phi )\over 2m} -\Phi^\dagger  V
\Phi  \right]
\label{2a}
\eeq where $D_\mu \Phi = \partial_\mu  \Phi  +\Phi A_\mu $.
Even though we have indicated the derivative as $\partial_\mu
\Phi$,  it should be emphasized that, since $\Phi$ is
noncommutative, even classically,  translations must be
implemented by taking commutators with an operator conjugate
to $x$. This is implicit in the definition of $\partial_\mu
\Phi$.  Further, in (\ref{2a}), the gauge fields act on the
right of the field $\Phi$ and the potential on the left. This
ensures that the action of the gauge field and the potential
commute and allows an unambiguous separation of these two
types of interaction terms. The one-particle wavefunction is
the matrix element of $\Phi$ between the vacuum and
one-particle states. The equation of motion for (\ref{2a}) is
\beq iD_0 \Phi + D_i (D_i \Phi ) - V \Phi  =0
\label{2b}
\eeq and taking the appropriate matrix element, we see that
the Schr\"odinger equation  has a similar form, with the
qualification that the action of derivatives is defined via
commutators with an operator conjugate to $x$.

With the algebra of the $x_i$'s in (\ref{1}), we can see that
translations of the argument of
$\Phi$ may be achieved using just the $x_i$'s themselves by
writing \cite{gross}
\beq -i \partial_i \Phi  = {1\over \theta} ~~[\epsilon_{ij}
x_j, \Phi] = {\epsilon_{ij} x_j \over
\theta} ~\Phi - \Phi ~{\epsilon_{ij} x_j \over \theta}
\label{3}
\eeq In other words, by using the adjoint action of
${\epsilon_{ij} x_j}$, we can obtain translations on
functions of $x_i$. It is then easy to check that
$[\partial_1, \partial_2]=0$. Translations thus involve the
left and right actions of the $x_i$'s on $\Phi$, which are
mutually commuting actions. Since we do not usually take
commutators of operators with the wavefunction in quantum
mechanics, it is preferable, in going to the one-particle
case, to replace the right action of the
$x$'s on $\Phi$ formally by a left action by
\beq x_{iR} \Phi = \Phi x_i\label{4}
\eeq This can be done in more detail as follows. If we
realize the $x$-algebra on a Hilbert space
${\cal H}$, then
$\Phi$ is an element of ${\cal H}\otimes {\tilde {\cal H}}$,
where ${\tilde {\cal H}}$ is the dual Hilbert space and we
can write
$\Phi=
\sum_{mn} \Phi_{mn} \vert m\ra \la n\vert$ in terms of a
basis $\{ \vert m\ra \}$. Mapping the elements of
${\tilde{\cal H}}$ to the corresponding elements of ${\cal
H}$ in the standard way, we can introduce $\Phi'=\sum_{mn}
\Phi_{mn} \vert m \ra \otimes \vert n\ra$. The right action
of $x$'s on $\Phi$ is then mapped onto the left action on
$\Phi'$ as given above. Note that $[x_{iR},x_{jR}] \Phi =
-\Phi [x_i,x_j]=-i\theta \epsilon_{ij} \Phi$. We can thus
identify
$-{\epsilon_{ij} x_{jR}}$ as
$k_i$ and we obtain the realization given in (\ref{2}). The
one-particle limit of field theory thus naturally leads to
the structure of two mutually commuting copies of the
$x$-algebra.

We see that from both points of view, namely, of one-particle
quantum mechanics as defined by an irreducible representation
of the Heisenberg algebra generalized to include
noncommutativity of coordinates, or as defined by the
one-particle limit of field theory, we are led to the
algebraic structure (\ref{1}, \ref{2}). This result is
consistent with the discussion of quantum mechanics on the
noncommuting two-sphere given in reference \cite{vpn}. In
that case also, one had two mutually commuting copies of the
$x$-algebra, which was $SU(2)$. The momentum operator was
then constructed from the $SU(2)\times SU(2)$ algebra in a
way analogous to the realization (\ref{2}). The present
results for the plane may in fact be obtained, as we shall
see later, for a small neighborhood of the sphere, in the
limit of large radius. 

A concrete and simple example which illustrates the general
discussion so far is provided by the harmonic oscillator on
the noncommutative plane. It is not any more difficult to
solve the more general case of a charged particle in a
magnetic field (the Landau problem) with  a quadratic (or
oscillator) potential and so we shall treat this case below. 
The fact that we have a magnetic field $B$ can be
incorporated by modifying the commutation rule for the momenta
to $[p_1,p_2]=iB$. In other words, $B$ measures the
noncommutativity of the momenta. The interplay of $B$ and
$\theta$ can thus lead to some interesting behavior.
Denoting the position and momentum operators by $\xi_i,
~i=1,...,4$, $\xi =(x_1,x_2,p_1,p_2)$,  the commutation rules
are
\beqar [\xi_i,\xi_j]&&= iP_{ij}\nonumber\\ P&&=\left(
\matrix{0&\theta&1&0\cr -\theta&0&0&1\cr -1&0&0&B\cr
0&-1&-B&0\cr}\right)\label{5}
\eeqar The Hamiltonian for the oscillator with magnetic field
is
\beq H= {1\over 2}\left[ p_1^2 +p_2^2 +\omega^2 (x_1^2 +x_2^2
)\right]
\eeq It is obviously invariant under rotations in the plane.
The angular momentum, being the generator of these rotations,
takes the form
\beq L=\frac{1}{1-\theta B} \left[ x_1 p_2 - x_2 p_1 +
\frac{B}{2} (x_1^2 + x_2^2 ) +
\frac{\theta}{2} (p_1^2 + p_2^2 )\right]
\eeq We observe that it acquires $\theta$-dependent
corrections compared to the commutative case.

The algebra (\ref{5}) has many possible realizations. The
`minimal' one in terms of two independent sets of canonical
coordinates and momenta $({\bar x}_i , {\bar p}_i )$
satisfying standard Heisenberg commutation relations would be
\beqar &x_1 = {\bar x}_1, ~~~~~~~~~~~~~~~~~ &p_1 = {\bar p}_1 + B
{\bar x}_2 \nonumber\\ &x_2 = {\bar x}_2 + \theta {\bar
p_1}, ~~~~~~~~~  &p_2 = {\bar p}_2
\eeqar We prefer, however, to use a realization as close to
(\ref{2}) as possible to maintain contact with noncommutative
field theory. Using the realization (\ref{2}) for the
momenta, we find $[k_1,k_2]=i(B-(1/\theta ))$. Because of
this, the cases $B<1/\theta$ and $B>1/\theta$ should be
treated differently. Consider first the case $B<1/\theta$. In
this case, we can define
\beqar x_1 =l \alpha_1 ,~~~~~~~~~~~~~~~&&p_1= {1\over
l}\beta_1 +q \alpha_2\nonumber\\ x_2= l \beta_1
,~~~~~~~~~~~~~~~&&p_2= {1\over l}\alpha_1 -q \beta_2\label{6}
\eeqar where $l^2=\theta$ and $q^2 = (1/\theta )-B$.
$\alpha_i, \beta_i$ form a set of canonical variables, i.e.,
$[\alpha_i ,\beta_j] =i\delta_{ij}$. The Hamiltonian for the
oscillator with the magnetic field is given by
\beqar H&&= {1\over 2}\left[ p_1^2 +p_2^2 +\omega^2 (x_1^2
+x_2^2 )\right]\nonumber\\ &&= {1\over 2}\left[
\left(\omega^2 l^2 +{1\over l^2}\right) (\alpha_1^2
+\beta_1^2) +q^2(\alpha_2^2 +\beta_2^2) +{2q\over l}
(\alpha_1 \beta_2 + \alpha_2 \beta_1)\right]
\label{7}
\eeqar We can now make a Bogolyubov transformation on this by
expressing $\alpha_i, \beta_i$ in  terms of  a canonical set
$Q_i,P_i$ by writing
\beq
\left(\matrix{\alpha_1\cr
\alpha_2\cr
\beta_1\cr
\beta_2\cr}\right)= \cosh \lambda \left(\matrix{Q_1\cr Q_2\cr
P_1\cr P_2\cr}\right) ~+~ \sinh\lambda \left(\matrix{P_2\cr
P_1\cr Q_2\cr Q_1\cr}
\right)\label{8}
\eeq Choosing
\beq
\tanh 2\lambda = -~{2ql \over 1+ \omega^2 l^4 + q^2
l^2}\label{9}
\eeq the Hamiltonian (\ref{7}) becomes
\beq H= {1\over 2} \left[ \Omega_+ ~(P_1^2+Q_1^2)+\Omega_{-}
(P_2^2 +Q_2^2)\right]
\label{10}
\eeq where
\beq
\Omega_{\pm} = \half \sqrt{(\omega^2\theta-B)^2 +4\omega^2}
~\pm \half (\omega^2 \theta +B) \label{11}
\eeq 
Equation (\ref{10}) shows that the spectrum is given by
that of two harmonic oscillators of frequencies $\Omega_+$
and $\Omega_{-}$.

The case of $B>1/\theta$ can be treated in a similar way.
With $q^2 =B-(1/\theta )$, we can write
\beqar x_1 =l \alpha_1 ,~~~~~~~~~~~~~~~&&p_1= {1\over
l}\beta_1 +q \alpha_2\nonumber\\ x_2= l \beta_1
,~~~~~~~~~~~~~~~&&p_2= -{1\over l}\alpha_1 +q
\beta_2\label{12}
\eeqar In terms of the $\alpha_i, \beta_i$, the Hamiltonian
becomes
\beq H= {1\over 2}\left[ \left(\omega^2 l^2 +{1\over
l^2}\right) (\alpha_1^2 +\beta_1^2) +q^2(\alpha_2^2
+\beta_2^2) +{2q\over l} (\alpha_2 \beta_1 -\alpha_1 \beta_2
)\right]
\label{13}
\eeq The required Bogolyubov transformation is
\beq
\left(\matrix{\alpha_1\cr
\alpha_2\cr
\beta_1\cr
\beta_2\cr}\right)= \cos \lambda \left(\matrix{Q_1\cr Q_2\cr
P_1\cr P_2\cr}\right) ~+~ \sin\lambda \left(\matrix{P_2\cr
P_1\cr -Q_2\cr -Q_1\cr}
\right)\label{14}
\eeq The required choice of $\lambda$ is given by
\beq
\tan 2\lambda ={2ql\over 1+\omega^2 l^4 -q^2 l^2}
\label{14a}
\eeq
$H$ can then be written as in (\ref{10}) with
\beq
\Omega_{\pm} = \pm \half \sqrt{(\omega^2\theta-B)^2
+4\omega^2} ~+ \half (\omega^2 \theta +B) \label{15}
\eeq  We again have two oscillators of frequencies
$\Omega_{\pm}$. 

We see from the above results that there is a critical value
of the magnetic field or
$\theta$ given by $B\theta =1$. $\Omega_{-}$ vanishes upon
approaching this value from either side. The Hamiltonian is
independent of $P_2, Q_2$. Thus the number of states for
fixed energy will become unbounded, since all the states
generated by $P_2,Q_2$ are now degenerate. This large
degeneracy can also be seen from a semiclassical estimate of
the number of states for fixed energy.  Going back to
(\ref{5}), we see that $\det P= (1-B\theta )^2$. The phase
volume is thus given by
\beqar d\mu &&= {1\over \sqrt{\det P}} ~dx_1 dx_2 dp_1 dp_2
\nonumber\\ &&= {1\over \vert 1-\theta B \vert }~dx_1 dx_2
dp_1 dp_2 \label{16}
\eeqar Surfaces of equal energy in phase space are ellipsoids
defined by $E=\half (p_1^2 + p_2^2 +\omega^2 x_1^2 + \omega^2
x_2^2 )$. A semiclassical estimate of the number of states
with energy less that $E$ is given by the volume inside this
surface divided by $(2\pi)^2$. We obtain
\beq N = \frac{V}{(2\pi)^2} = {1 \over 2\vert 1-B\theta \vert
} \left(\frac{E}{\omega}\right)^2
\label{17}
\eeq The criticality of the point $\theta B =1$ is once again
clear; the density of states is infinite  at this point.

When $\omega^2 = 0$ we have the pure Landau problem. In this
case $\Omega_+  = B,~ \Omega_- =0$ for $B>0$, or $\Omega_+
=0,~ \Omega_- = |B|$ for $B<0$ and we have the standard,
infinitely degenerate Landau levels as in the commutative
case. The density of states per unit area, denoted by $\rho$,
however, is now modified to 
\beq
\rho = \frac{1}{2\pi} \left| \frac{B}{1-\theta B} \right|
\label{rho}\eeq To demonstrate this, observe that the
magnetic translations, defined as the operators performing
translations on $x_i$ and commuting with the Hamiltonian, are
now
\beq D_1 = \frac{1}{1-\theta B}(p_1 - B x_2) ,~~~~ D_2 =
\frac{1}{1-\theta B}(p_2 + B x_1)
\eeq These are the operators responsible for the infinite
degeneracy of the Landau levels and their commutator
determines the density of degenerate states on the plane.
$D_i$ commute with $x_j$ in the standard way,
$[x_i , D_j ] = i \delta_{ij}$, but their mutual commutator
is now
\beq [ D_1 , D_2 ] = - i \frac{B}{1-\theta B} \label{DD}
\eeq which reproduces the result (\ref{rho}). We observe that
for the critical value of the magnetic field $B=1/\theta$ the
density of states on the plane becomes infinite.  The same
result can also be obtained in the semiclassical way of the
previous paragraph, where now we calculate the phase space
volume of a circle 
$E=\half (p_1^2 + p_2^2 )$ in momentum space times a domain
of area $A$ in coordinate space. The result is
\beq N = \frac{V}{(2\pi)^2} = \frac{E A}{2\pi |1-\theta B|}
\eeq which is compatible with (\ref{rho})  upon filling the
lowest $n$ Landau levels such as $E=n |B|$.

It is also interesting to calculate the magnetic length in
this case, that is, the minimum spatial extent of a
wavefunction in the lowest Landau level. This can be achieved
by putting both oscillators $\Omega_+$ and $\Omega_-$ in
their ground state: the one with nonvanishing frequency
excites Landau levels while the one with vanishing frequency
creates annular states on the plane for each Landau level. In
this state we have $\la P_i^2 \ra = \la Q_i^2 
\ra = \half $ and $\la P_i \ra = \la Q_i \ra =0$. Using
(\ref{6}), (\ref{8}) and (\ref{9}) we can calculate $\la
x_i^2 \ra$ for $B<1/\theta$ as
\beq
\la x_1^2 + x_2^2 \ra = l^2 (\cosh^2 \lambda + \sinh^2
\lambda ) = \frac{1}{|B|}\left( 2-B\theta \right)
\eeq while for $B > 1/\theta$ we obtain from (\ref{12}) and
(\ref{14})
\beq
\la x_1^2 + x_2^2 \ra = l^2 (\cos^2 \lambda + \sin^2 \lambda
) = \theta
\eeq So we see that for subcritical magnetic field the
magnetic length is more or less as in the commutative case
while for overcritical one it assumes the value $l =
\sqrt{\theta}$ which is the minimal uncertainty on the
noncommutative plane.

We conclude by noting that the oscillator frequency $\omega$
and magnetic field $B$ appearing in the Hamiltonian are
distinct from the corresponding `kinematical' quantities that
appear in the equations of motion. Expressing the equations
of motion in terms of
$x_i$ and its time derivatives we obtain
\beq {\ddot x}_i = (B+\theta \omega^2 ) \epsilon_{ij} {\dot
x}_i - (1-\theta B) \omega^2 x_i
\eeq We recognize the Lorenz force and the spring force with
effective magnetic field and spring constant
\beq {\tilde B} = B+\theta \omega^2 ,~~~~ {\tilde \omega}^2 =
(1-\theta B) \omega^2
\eeq The spectral frequencies $\Omega_\pm$ in terms of the
kinematical parameters become identical to the corresponding
noncommutative ones, namely
\beq
\Omega_\pm = \left| \pm \half \sqrt{{\tilde B}^2 + 4{\tilde
\omega}^2} +\half {\tilde B} \right|
\eeq In this parametrization the noncommutativity of space
manifests only through the density of states and spatial
correlation functions. Interestingly, for the critical value
$B=1/\theta$ the oscillator $\omega$ transmutates entirely
into a magnetic field
${\tilde B} = \theta \omega^2 + \theta^{-1}$

\vskip .1in
\section{The noncommutative sphere}
\vskip .1in We now turn to the quantum mechanics of a
particle on the noncommutative two-sphere. To set the stage,
we first give a review of the commutative sphere with a
magnetic monopole at the center.  The observables of the
theory consist of the particle coordinates $x_i$ and the
angular momentum generators $J_i$, $i=1,2,3$. Their algebra is
\beqar {[} x_i , x_j {]} &=&0 \nonumber\\ {[} J_i , x_j {]}
&=& i \epsilon_{ijk} x_k \nonumber\\ {[} J_i , J_j {]} &=& i
\epsilon_{ijk} J_k \label{xJ1}
\eeqar while the Hamiltonian is taken to be
\beq H = \frac{1}{2 x^2} J^2 \label{HC}
\eeq The algebra (\ref{xJ1}) has two Casimirs, which can be
chosen to have fixed values, say, 
\beqar x^2 &=& a^2\nonumber \\ x \cdot J &=& -\frac{n}{2} a
\eeqar The first one is simply the square of the radius of
the sphere. In the second one, $n$ can be  identified as the
monopole number. Indeed, in the presence of a monopole field
the angular momentum acquires a term in the radial direction
proportional to the monopole number which makes the second
Casimir nonvanishing. The interpretation as a magnetic field
can be independently justified by deriving the equations of
motion of $x_i$ using the Hamiltonian (\ref{HC})
\beq {\ddot x}_i =~-~ \frac{1}{2} \left[ \left(\frac{J^2 -
(n/2)^2}{a^4} \right) x_i + x_i
\left(\frac{J^2 - (n/2)^2}{a^4} \right) \right] +
\epsilon_{ijk} \, {\dot x}_j \,
\frac{n x_k}{2 a^3}
\eeq The first term is a centripetal force, due to the motion
on a curved manifold; the kinematical angular momentum
squared is seen to be $J^2 - (n/2)^2$. The second term is a
Lorentz force, corresponding to a radial magnetic field $B_i
= (n/2) {\hat x}_i /a^2$. The monopole number, then, is
\beq N = \frac{4\pi a^2 B}{2\pi} = n
\eeq It is interesting that the magnetic field does not
appear as a parameter in the Hamiltonian, not even as a
modification of the Poisson structure (as in the planar
case), but rather as a Casimir of the algebra of observables.

We now turn to the noncommutative sphere. The quantum
mechanics of a particle on a noncommutative sphere was
discussed in \cite{vpn}. The structure of observables is
similar, with the difference that the coordinates do not
commute but rather form an 
$SU(2)$ algebra. Specifically,
\beqar {[} R_i , R_j {]} &=& i \epsilon_{ijk} R_k \nonumber\\
{[} J_i , R_j {]} &=& i \epsilon_{ijk} R_k \nonumber\\ {[}
J_i , J_j {]} &=& i \epsilon_{ijk} J_k \label{xJ2}
\eeqar
$R^2$ is a Casimir, as before, but the magnetic Casimir is
deformed to $R \cdot J - \half J^2$.

This operator structure is realized in terms of an
$SU(2)\times SU(2)$-algebra with corresponding mutually
commuting generators $R_i$, $K_i$. In terms of these, the
angular momentum is $J_i =R_i +K_i$. We have two Casimir
operators, $R^2 =r(r+1)$ and $K^2=k(k+1)$, and we can label
an irreducible representation by the maximal spin values
$(r,k)$. The magnetic Casimir becomes
$\half (R^2 - K^2 )$. If the radius of the sphere is denoted
by $a$ as before,  we can identify the  coordinates $x_i$ as
\beq x_i = {a\over \sqrt{r(r+1)}}~R_i
\label{18}
\eeq The commutative case can be obtained as the limit in
which both $r$ and $k$ become very large, but with their
difference $k-r=n/2$ being fixed (so that the angular
momentum 
$J^2$ remain finite). In that limit, the magnetic Casimir
becomes
\beq
\half (R^2 - K^2 ) = (r-k) \frac{k+r+1}{2} \simeq
-\frac{n}{2} r
\eeq
$n$ becomes the monopole number. We can, therefore, identify
the integer $n=2(k-r)$ as a quantized `monopole' number in
the noncommutative case. 

The Hamiltonian of the particle can again be taken
proportional to the square of the angular momentum:
\beq H = \frac{\gamma}{2 a^2} J^2 \label{Hgamma}
\eeq with $\gamma$ some coefficient depending on the
Casimirs. In the limit of a commutative sphere $\gamma$
should become $1$ in order to reproduce the standard results.
In general, however, there is no a priori reason to fix a
specific value for $\gamma$ and, as we shall demonstrate, a
different choice must be made in order to recover the limit
of the noncommutative plane. The energy spectrum of the
particle is clearly
\beq E = \frac{\gamma}{2 R^2} j(j+1) ~,~~~~ j =
\frac{|n|}{2},  \frac{|n|}{2}+1 ,\dots j+k
\eeq Both the energy and angular momentum have a finite
spectrum, reflecting the fact that the Hilbert space is
finite dimensional.

Comparison to the noncommutative plane can be made by scaling
appropriately the parameters of the model. We should take
the radius $a$ in (\ref{18}) to infinity and consider a small
neighborhood, say, around the `north pole' $R_3=r$, with
$x_1,x_2$ being  the relevant coordinates. {}From the
definition (\ref{18}) of $x_i$, we then identify the
noncommutativity parameter as
$\theta\approx a^2/r$. So the scaling of the parameter $r$ is
\beq r = \frac{a^2}{\theta} ~,~~~a \to \infty \label{rR}
\eeq Only the low-lying states of $J^2$ and $H$ should be
considered in this limit, with $j=
\frac{|n|}{2}+l$, $l=0,1,2\dots$. Since $R_3 \approx r$ for
the states of interest in this limit, we must also have $K_3
\approx -k = -r - n/2$ so that $j \approx |n|/2$. This also
means that $J_3 \approx  -n/2$. $k=r+n/2$ and $\gamma$ should
then scale  appropriately to obtain the planar operator
algebra of  observables for a particle on the  noncommutative
plane in the presence of a magnetic field.

The operators $\epsilon_{ij} J_j /a$ ($i,j=1,2$) generate
translations of $x_i$ in the planar limit, i.e.,
\beq [ x_i , \frac{1}{a} \epsilon_{jk} J_k ] =
\frac{i}{\sqrt{r(r+1)}} \delta_{ij} R_3 \approx i
\delta_{ij}
\eeq So one might be tempted to identify them with the
momentum operators in that limit. In the presence of a
magnetic field, however, we understand that these should
instead become the magnetic translations $D_i$, since they
both commute with the Hamiltonian. Their commutator
\beq [ D_1, D_2 ] = \frac{1}{a^2} [ J_2 , -J_1 ] =
\frac{i}{a^2} J_3 \approx -i \frac{n}{2 a^2}
\label{DDJ}\eeq should then reproduce the result (\ref{DD})
for the plane. This leads to the identification
\beq n = \frac{2B a^2}{1-\theta B} \label{nB}
\eeq which fixes the scaling of $n$ and $k$. It remains to
identify the momenta $p_i = {\dot x}_i$. {}From the
Hamiltonian (\ref{Hgamma}) we obtain
\beq {\dot x}_i = \frac{\gamma}{R\sqrt{r(r+1)}}
\epsilon_{ijk} K_j R_k
\eeq The commutator of $x_i$ and $p_j = {\dot x}_j$ then
becomes
\beq [ x_i , p_j ] = \frac{i \gamma}{r(r+1)} \left( K_i R_j -
K_k R_k \delta_{ij} \right)
\eeq In the planar limit $K_3$ and $R_3$ dominate over
$K_{1,2}$ and $R_{1,2}$. Therefore, the above commutator
becomes, for $i,j=1,2$,
\beq [ x_i , p_j ] \approx  - \frac{i \gamma}{r^2} K_3 R_3
\delta_{ij} \approx i \gamma  
\frac{k}{r} \delta_{ij}
\eeq (we also set $r(r+1) \approx r^2$). To reproduce the
canonical commutators on the plane we must set
\beq
\gamma = \frac{r}{k} = \frac{r}{r+\frac{n}{2}} = 1-\theta B
\eeq which fixes the scaling of $\gamma$. We can now
calculate the commutator of momenta
\beq [ p_1 , p_2 ] = \frac{i}{k^2 a^2} K \cdot R (K_3 + R_3)
\approx i \frac{n r}{2 a^2 k} = iB
\eeq which is, indeed, the correct planar commutator. 

Finally, the spectrum of the Hamiltonian becomes
\beq E = \frac{\gamma}{2a^2} \left( \frac{|n|}{2} +l\right)
\left( \frac{|n|}{2} +l +1\right)
\approx \frac{\gamma n^2}{8a^2} + \frac{\gamma |n|}{2a^2} (l
+ \half ) = 
\frac{B^2 a^2}{2(1-\theta B)} + |B| (l+\half )
\eeq Apart from a zero-point shift of order $a^2$, we have
agreement with the Landau level spectrum of the
noncommutative plane. The above spectrum, but without the
zero-point shift, is also reproduced by the low-lying states
of the operator $H' = \half p_i^2$, thus establishing the
full correspondence with the plane. The density of states on
each Landau level can also be calculated. For a given energy
eigenvalue corresponding to $j=|n|/2 + l$ there are $2j+1$
degenerate states. The space density of these states is
\beq
\rho = \frac{2j+1}{4\pi a^2} = \frac{|n|}{4\pi a^2} +
\frac{2l+1}{4\pi a^2} \approx
\frac{1}{2\pi} \left| \frac{B}{1-\theta B} \right|
\eeq again in agreement with the planar case.

A final word is in order concerning the sphere-plane
correspondence. Naively, we may expect that to obtain the
full range of magnetic fields $-\infty < B < +\infty$, the
integer $n$ should span all values from $-\infty$ to
$+\infty$. But clearly this is not possible,  since $k=r+n/2
=r(1+\mu )$ must be nonnegative. In other words, $\mu =n/2r$
must satisfy $\mu >-1$. {}From (\ref{rR},
\ref{nB}) we have
\beq
\mu = \frac{\theta B}{1-\theta B}
\eeq We see that the allowed values of $\mu$ correspond to
$B<1/\theta$. $B \to 1/\theta$ corresponds to $\mu \to
\infty$ and $B \to -\infty$ corresponds  to $\mu \to 1$. What
about
$B > 1/\theta$? For these fields we observe that $\gamma =
1-\theta B <0$ and thus the coefficient of $J^2$ in the
Hamiltonian (\ref{Hgamma}) becomes negative. This means that
low-lying energy states now correspond to the {\it highest}
value of $j$ (rather than the lowest), that is,
$j=r+k=2r+n/2$. Since $R_3 \approx r$, we must have $K_3
\approx k$ for such states, and thus $J_3 \approx r+k$.
Repeating the previous analysis, we see from (\ref{DDJ}) that
it is $-J_3 /R^2$ which is identified with $B/(1-\theta B)$
and thus
\beq r+k = -\frac{B R^2}{1-\theta B} ~~~~ {\rm so~ that}~~~~
\mu =\frac{n}{2r}  = \frac{2-\theta B}{\theta B -1}
\eeq For $B>1/\theta$ the above $\mu$ spans again the allowed
range of values $q>-1$. In summary, the picture is that for
each pair of values $r,k$ the energy spectrum of the sphere
is bounded between two end levels, one at $j_- =|r-k|$ and
one  at $j_+ =r+k$. The spectrum around each end level maps
to a particular magnetic field in the planar limit, $j_-$
corresponding to subcritical and $j_+$ to overcritical values
of $B$. Thus the mapping from sphere to plane is actually one
to two. This also suggests a duality between  magnetic fields
$B$ and $B' $ satisfying $B + B' = 2/\theta$, since they both
correspond to the same spherical case. The critical field $B
= 1/\theta$ is self-dual.

\vskip .2in
\leftline{\bf Acknowledgments}

This work was supported in part by the National Science
Foundation grant PHY-0070883, a PSC-CUNY-31 award and a CUNY
Collaborative Incentive Research Grant.

\end{document}